\documentclass[a4paper,10pt]{article}
\usepackage[utf8x]{inputenc}
\usepackage{amsmath,amssymb,graphics,graphicx,dcolumn,bm,enumerate}

\title{Econophysics Research  in India in  the last two Decades}
\author{Asim Ghosh \\ Theoretical Condensed Matter Physics Division \\ Saha Institute of Nuclear Physics \\ 1/AF Bidhannagar, Kolkata 700 064, India.}

\date{}
\begin{document}

\maketitle

\begin{abstract}
\noindent We discuss here researches  on econophysics done from India in the last two decades. The term `econophysics' was formally coined  in India (Kolkata) in 1995. Since then  many research papers, books, reviews,  etc.  have been written by  scientists. Many institutions are now  involved in  this research field and many  conferences are being  organized there.  In this article we give an account (of papers, books, reviews, papers in proceedings volumes etc.) of  this research from India. 
\end{abstract}

\section{Introduction}
\noindent The subject econophysics is an interdisciplinary research field where the tools of physics are applied to understand the problem of economics. The term `econophysics' was coined  by   H. Eugene Stanley  in a Kolkata conference on statistical physics in $1995$.  The research on economics by physicists is not  new. There were many physicists who contributed significantly in the  development of  economics. For example  Daniel Bernoulli,  who developed utility-based preferences,  was a physicist. Similarly  Irving Fisher, who was one of the founders of neoclassical economic theory, was a student of statistical physicist Josiah Willard Gibbs. Also Jan Tinbergen, who won the first Nobel Prize in economics,   did his Ph. D.  in statistical physics in Leiden university under Paul Ehrenfest. However these physicists (by training) eventually left physics and migrated to economics. The new feature of the developments for the last two decades is that physicists studying the problems of economics or sociology  remain in their respective  departments  and publish their econophysics research results in almost all the major  physics journals.  

In India,  works of such interdisciplinary nature   are not  new.  The Indian Statistical Institute, Kolkata,  is the one of oldest institutions in India (founded in 1931). The main motivation of this research institute was to promote interactions of  natural and social sciences;  in particular  to advance the role of statistics. The work on econophysics in India  started around 1990 from  Saha Institute Nuclear Physics, Kolkata. Now-a-days many researchers from different universities and   institutes from our country are also involved in this research field and  international conferences  are being  organized here on regular basis.

Over the last two decades  many papers, books and reviews have been written by the Indian scientists in this field. We will analyze  here  the statistics of such publications and other endeavors.

\section{A statistical survey on the development of econophysics [world wide]}
\begin{figure}[h]
\centering
\includegraphics[height=6.cm]{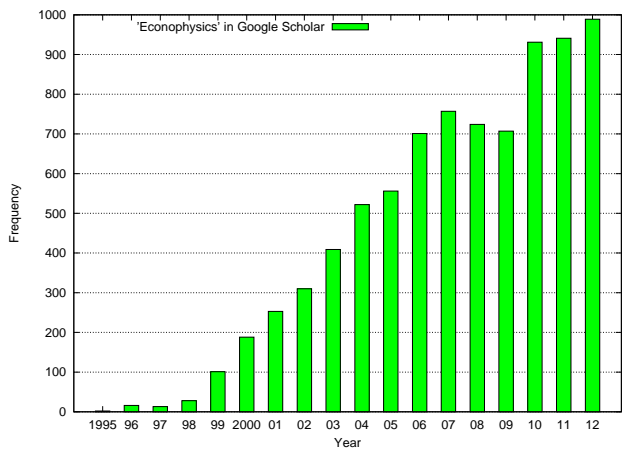}
\caption{Histogram plot of  numbers of entries containing the term `econophysics' versus the corresponding year. The   data are taken from google scholar site (http://scholar.google.co.in/schhp).}
\label{fig1}
\end{figure}

To see how the subject grew after introduction of this topic (term)  in 1995  in scientific community all over the  world, we have taken  the statistics of the articles having  `econophysics' term  any where in the articles from   google scholar site. Fig. \ref{fig1} shows histogram plot of the number of papers posted in google scholar over different years. The figure  clearly indicates  that the subject is growing quite fast, starting in around 1995.  

\begin{figure}[h]
\centering
\includegraphics[height=6.cm]{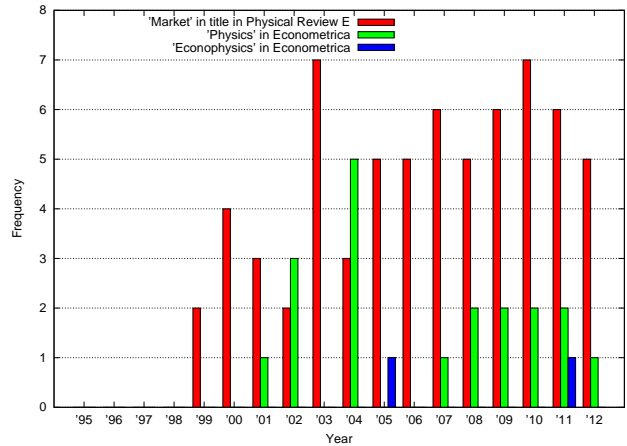}
\caption{Histogram plot of  numbers of papers containing the term `Market' in the title of the journal `Physical Review E', the terms  `physics' and `econophysics' in the journal `Econometrica' (data from respective journal website).}
\label{fig2}
\end{figure}

To get  an idea about the impact of econophysics research on physics as well as  economics,  we give in Fig. \ref{fig2}, the count of the papers   having  the term `market' in the titles of a typical statistical  physics journal, namely  `Physical Review E' (published by American Physical Society) and  the  number of  papers having the term `physics' or  `econophysics' in a typical  economics journal `Econometrica' (published by Econometric Society) for the same period.

\section{Books published  from India }
We give below a list of books written by Indian scientists \cite{book-sinha,book-Guhathakurta,book-Chakrabarti,book-sen}:

\begin{itemize}
\item S. Sinha, A. Chatterjee, A. Chakraborti and B. K. Chakrabarti, \textit{Econophysics:An Introduction}, Wiley-VCH, Berlin, 2010. \newline 
[Book Chapters: \textbullet~ Introduction \textbullet~ The Random Walk\textbullet~ Beyond the Simple Random Walk\textbullet~Understanding Interactions through Cross-Correlations \textbullet~Why care about a Power Law?  \textbullet~The Log-Normal Distribution \textbullet~When a Single Distribution is not Enough  \textbullet~ Explaining Complex Distributions with Simple Models \textbullet~But Individuals are not Gas Molecules\dots  \textbullet~...and Individuals don't Interact Randomly: Complex Networks \textbullet~Outlook and Concluding Thoughts  \textbullet~Appendices: \textbullet~Thermodynamics and Free Particle Statistics  \textbullet~Interacting Systems: Mean Field Models, Fluctuations and Scaling Theories \textbullet~Renormalization Group Technique \textbullet~Spin Glasses and Optimization Problems: Annealing \textbullet~Nonequilibrium Phenomena.]

\item  K. Guhathakurta, B. Bhattacharya and A. Roychowdhury,   \textit{Examining Stock Markets : a non linear dynamics perspective: Examining the Geometric Brownian Motion model with respect to Stock Price Movement in an Emerging Market}, LAP LAMBERT Academic Publishing, 2012. \newline 
[Book Chapters: \textbullet~ Introduction \textbullet~ Related literature  \textbullet~ Theoretical framework of the models under study \textbullet~ Research methodology \textbullet~ Empirical mode decomposition analysis of financial time series  \textbullet~ Recurrence analysis of critical regimes of stock market \textbullet~ Examining the geometric brownian motion and comparison with borland model \textbullet~ Concluding observations.]

\item B. K. Chakrabarti, A. Chakraborti, S. R. Chakravarty and A. Chatterjee, \textit{Econophysics of Income and Wealth Distributions}, Cambridge University Press, Cambridge, 2013.\newline 
[Book Chapters: \textbullet~Introduction \textbullet~Income and wealth distribution data for different countries \textbullet~Major socio-economic modellings \textbullet~Market exchanges and scattering process \textbullet~ Analytic structure of the kinetic exchange market models \textbullet~Microeconomic foundation of the kinetic exchange models  \textbullet~Dynamics: generation of income, inequality and development \textbullet~Outlook.]

\item P. Sen,  B. K. Chakrabarti,\textit{ Sociophysics: An introduction}, Oxford University Press, Oxford, 2013. \newline 
[Book Chapters: \textbullet~Introduction \textbullet~Basic features of social systems and modelling \textbullet~ Opinion formation in a society  \textbullet~Social choices and popularity \textbullet~Crowd avoiding dynamical phenomena \textbullet~Social phenomena on complex networks \textbullet~Of flocks, flows and transports \textbullet~  Endnote \textbullet~Appendices: \textbullet~Phase transitions and critical phenomena \textbullet~ Magnetic systems: static and dynamical behaviour \textbullet~Percolation and fractals \textbullet~Random walks \textbullet~ Monte Carlo simulations \textbullet~Some data analysis methods and useful tables.]

\end{itemize}


\section{Papers published from India}
 Here we  give the list of papers published by Indian  scientists in international (refereed) journals  from 1995-till date (the Indian cities, where the work was done, are indicated in  the third bracket).      
\vskip 0.2cm
\noindent \textbf{1995}
\noindent \textbullet~B. K. Chakrabarti and S. Marjit, \textit{Self-organisation in Game of Life and economics},  Indian J. Phys. B \textbf{69 } 681 (1995) [Kolkata].

\vskip 0.2cm

\noindent\textbf{2000}
\textbullet~A. Chakraborti and B. K. Chakrabarti,  \textit{Statistical mechanics of money: how saving propensity affects its distribution}, Eur. Phys. J., \textbf{B 17} 167 (2000) [Kolkata].

\vskip 0.2cm

\noindent  \textbf{2001}
\textbullet~
A. Chakraborti, S. Pradhan and B. K. Chakrabarti,  \textit{A self-organising model of market with single commodity}, Physica A \textbf{297} 253 (2001) [Kolkata].

\vskip 0.2cm
\noindent \textbf{2003}
\noindent  \textbullet ~ S. Sinha, \textit{Stochastic maps, wealth distribution in random asset exchange models and the marginal utility of relative wealth}, Phys. Scripta T \textbf{106} 59 (2003) [Chennai]. \textbullet ~ A. Das and S. Yarlagadda,\textit{ Analytic treatment of a trading market model}, Phys. Scripta T \textbf{ 106} 39 (2003) [Kolkata]. \textbullet~ A. Chatterjee, B. K. Chakrabarti and S. S. Manna,  \textit{Money in gas-like markets: Gibbs and Pareto laws}, Phys. Scripta  T \textbf{106} 36 (2003) [Kolkata].

\vskip 0.2cm

\noindent \textbf{2004}
\textbullet~ A. Chatterjee, B. K. Chakrabarti and S. S. Manna,  \textit{Pareto law in a kinetic model of market with random saving propensity}, Physica A \textbf{335} 155 (2004) [Kolkata]. \textbullet~ S. Sinha and S. Raghavendra, \textit{Hollywood blockbusters and long-tailed distributions: An empirical study of the popularity of movies}, Eur. Phys. J. B \textbf{42} 293 (2004) [Chennai]. 
\vskip 0.2cm

\noindent \textbf{2005}
\textbullet~A. Chatterjee, B. K. Chakrabarti and R. B. Stinchcombe, \textit{Master equation for a kinetic model of trading market and its analytic solution},  Phys. Rev.  E \textbf{72} 026126 (2005) [Kolkata \& Oxford].
\textbullet~ A. Chakraborti and M. S. Santhanam, \textit{Financial and other multivariate time series : spectral and related properties},  Int. J. Mod. Phys. C \textbf{16} 1733 (2005) [New York \& Ahmedabad]. 
\textbullet~
P. Manimaran, P. K. Panigrahi, and J. C. Parikh,  \textit{Wavelet analysis and scaling properties of time series},Phys. Rev. E \textbf{72} 046120 (2005) [Hyderabad \& Ahmedabad] 

\vskip 0.2cm

\noindent \textbf{2006}
\textbullet~
S. Sinha, \textit{Evidence for power-law tail of the wealth distribution in India},  Physica A \textbf{359} 555 (2006) [Chennai].
\textbullet~
A. Chatterjee and B. K. Chakrabarti, \textit{Kinetic market models with single commodity having price fluctuations},  Eur. Phys. J. B \textbf{ 54} 399 (2006) [Kolkata].
\textbullet~
P. K. Mohanty, \textit{Generic features of the wealth distribution in ideal-gas-like markets}, Phys. Rev.  E \textbf{74} 011117 (2006) [Kolkata].
\textbullet~ 
A. Kargupta, \textit{Money exchange model and a general outlook},  Physica A  \textbf{359} 634 (2006) [Panskura]. 
\textbullet~ M. S. Santhanam, J. Bandyopadhyay and D. Angom,  \textit{Quantum spectrum as a time series : Fluctuations and self-similarity}, Phys. Rev. E  \textbf{73} 015201(R) (2006) [Ahmedabad]. 
\textbullet~ D. P. Ahalpara and   J. C. Parikh, \textit{Modeling time  series  data of the real systems},  Int. J. Mod. Phys. C \textbf{18} 235 (2007) [Ahmedabad].

\vskip 0.2cm

\noindent \textbf{2007}
\textbullet~A. Chatterjee and B. K. Chakrabarti,   \textit{Kinetic Exchange Models for Income and Wealth Distributions}, Eur. Phys. J. B \textbf{ 60} 135 (2007) [Kolkata].
\textbullet~ A. Chatterjee, S. Sinha and B. K. Chakrabarti, \textit{Economic inequality: Is it natural?},  Current Science \textbf{92} 1383 (2007) [Kolkata \& Chennai]. \textbullet~  V. Kulkarni and  N. Deo,  \textit{ Correlation and volatility in an Indian stock market: A random matrix approach}, Eur. Phys. J. B \textbf{60 }  101 (2007) [Delhi]. 
\textbullet~
R. K. Pan and S. Sinha, \textit{Collective behavior of stock price movements in an emerging market}, Phys. Rev. E \textbf{76}  046116 (2007) [Chennai].
\textbullet~
R. K. Pan and S. Sinha, \textit{Self-organization of price fluctuation distribution in evolving markets}, Europhys. Lett. \textbf{77} 58004(2007) [Chennai]. 
\textbullet~
A. Chatterjee, B. K. Chakrabarti, \textit{Ideal-gas like market models with savings: quenched and annealed cases}, Physica A \textbf{382} 36 (2007) [Kolkata].
\textbullet~
P. Bhattacharyya, A. Chatterjee , B. K. Chakrabarti, \textit{A common origin of the power law distributions in models of market and earthquake}, Physica A \textbf{381} 377 (2007) [Kolkata]. 
\textbullet~ M. A. Saif and P. M. Gade, \textit{Emergence of Power Law in a Market with Mixed Models},  Physica A, \textbf{384} 448 (2007) [Pune].

\vskip 0.2cm

\noindent \textbf{2008}
\textbullet~
U. Basu and P. K. Mohanty, \textit{Modeling wealth distribution in growing markets},  Eur. Phys. J. \textbf{B 65} 585 (2008) [Kolkata].
\textbullet~A. Kargupta, \textit{Relaxation in the wealth exchange models}, Physica A \textbf{387} 6819 (2008) [Panskura].
\textbullet~ R. K. Pan and S. Sinha, \textit{Inverse-cubic law of index fluctuation distribution in Indian markets}, Physica A \textbf{387}   2055(2008) [Chennai]. 
\textbullet~ K. Guhathakurta, M. Indranil, and A. C. Roy,  \textit{Empirical mode decomposition analysis of two different financial time series and their comparison Chaos}, Solitons \& Fractals \textbf{37}  1214 (2008) [Kolkata].

\vskip 0.2cm

\noindent \textbf{2009}
\textbullet~  
A. S. Chakrabarti and B. K. Chakrabarti,  \textit{Microeconomics of the ideal gas like market model} Physica A \textbf{388} 4151 (2009) [Kolkata].
\textbullet~ 
A. S. Chakrabarti, B. K. Chakrabarti, A. Chatterjee and M. Mitra, \textit{The Kolkata Paise Restaurant Problem and resource utilization}, Physica A \textbf{388}  2420 (2009) [Kolkata \& Trieste]. 
\textbullet~ 
S. Kumar, and  N. Deo,\textit{ Multifractal properties of the Indian financial market}, Physica A  \textbf{388} 1593 (2009) [Delhi].  
\textbullet~ K. Gangopadhyay and  B. Basu, \textit{City Size Distributions For India and China}, Physica A  \textbf{388} 2682 (2009) [Kolkata].
\textbullet~ S. Biswas and P. Sen, \textit{Model of binary opinion dynamics: coarsening and effect of disorder},  Phys. Rev. E \textbf{80} 027101 (2009) [Kolkata]. 
\textbullet~ B. Basu, and S. Bandyapadhyay, \textit{Zipf's law and distribution of population in Indian cities} Indian Journal of Physics \textbf{83.11} 1575(2009) [Kolkata].
\textbullet~
M. A. Saif and P. M. Gade, \textit{Effects of Introduction of New Resources and Fragmentation of Existing Resources on Limiting Wealth Distribution in Asset Exchange Models}, Physica A \textbf{388}, 697 (2009) [Pune].

\vskip 0.2cm
\noindent\textbf{2010}
\textbullet~
A. S. Chakrabarti and B. K. Chakrabarti, \textit{Statistical theories of income and wealth distribution}, Economics E-journal, \textbf{4} (2010) [Kolkata].
\textbullet~
A. S. Chakrabarti and B. K. Chakrabarti,  \textit{Inequality reversal: Effects of savings propensity and correlated returns}, Physica \textbf{A 389} 3572 (2010) [Kolkata].
\textbullet~ A. Ghosh, A. Chatterjee, Manipushpak Mitra and B. K. Chakrabarti, \textit{Statistics of the Kolkata Paise Restaurant Problem}, New J. Phys. {\bf 12} 075033 (2010) [Kolkata \& Trieste].
\textbullet~ A. Chatterjee and P. Sen, \textit{Agent dynamics in kinetic models of wealth exchange}, Phys. Rev. E \textbf{82} 056117 (2010) [Kolkata \& Trieste].
\textbullet~ R. K. Pan and S. Sinha, \textit{The statistical laws of popularity: Universal properties of the box office dynamics of motion pictures}, New  J. Phys. \textbf{12}  115004(2010) [Chennai].
\textbullet~ M. Lallouache, A. S. Chakrabarti, A. Chakraborti and B. K. Chakrabarti, \textit{Opinion formation in kinetic exchange models: Spontaneous symmetry-breaking transition},  Phys. Rev. E \textbf{82 }056112 (2010) [Paris, Boston \&  Kolkata].
\textbullet~
K. Guhathakurta, B. Bhattacharya and A. Roy Chowdhury,  \textit{Using recurrence plot analysis to distinguish between endogenous and exogenous stock market crashes},   Physica A \textbf{389 } 1874 (2010) [Kolkata].

\vskip 0.2cm
\noindent \textbf{2011}
\textbullet~ D. Dhar, V. Sasidevan, B. K. Chakrabarti,  \textit{ Emergent cooperation amongst competing agents in minority games} Physica A \textbf{390} 3477 (2011) [Mumbai].
\textbullet~ S. Goswami, A. Chatterjee and P. Sen,  \textit{Antipersistent dynamics in kinetic models of wealth exchange}, Phys. Rev. E \textbf{84} 051118 (2011) [Kolkata].
\textbullet~
A. Ghosh, U. Basu, A. Chakraborti, B. K. Chakrabarti, \textit{Threshold-induced phase transition in kinetic exchange models}, Phys. Rev. E \textbf{83} 061130 (2011) [Kolkata \& Paris].
\textbullet~
S. V. Vikram and S. Sinha,  \textit{Emergence of universal scaling in financial markets from mean-field dynamics}, Phys. Rev. E \textbf{83} 016101(2011) [Chennai].
\textbullet~
S. Biswas, \textit{Mean-field solutions of kinetic-exchange opinion models},  Phys. Rev. E \textbf{84} 056106 (2011) [Kolkata].
\textbullet~
P. Sen, \textit{Phase transitions in a two-parameter model of opinion dynamics with random kinetic exchanges},  Phys. Rev. E \textbf{83} 016108 (2011) [Kolkata]. 
\textbullet~
S. Biswas, A. K. Chandra, A. Chatterjee and B. K. Chakrabarti, P\textit{hase transitions and non-equilibrium relaxation in kinetic models of opinion formation}, J.  Phys.: Conf. Ser.  \textbf{297} 012004 (2011) [Kolkata \& Marseille]. 
\textbullet~A. Ghosh, K. Gangopadhyay, and B. Basu,  \textit{Consumer expenditure distribution in India, 1983–2007: Evidence of a long Pareto tail},  Physica A \textbf{ 390} 83 (2011) [Kolkata \& Kozhikode].

\vskip 0.2cm
\noindent \textbf{2012}
\textbullet~ 
S. Biswas, A. Ghosh, A. Chatterjee, T. Naskar, B. K. Chakrabarti,  \textit{Continuous transition of social efficiencies in the stochastic-strategy minority game}, Phys. Rev. E \textbf{85} 031104 (2012) [Kolkata \& Marseille].
\textbullet~ 
A. Ghosh, D. D. Martino, A. Chatterjee, M. Marsili, B. K. Chakrabarti, \textit{ Phase transitions in crowd dynamics of resource allocation}, Phys. Rev. E {\bf 85}, 021116 (2012) [Kolkata, Marseille, Trieste \& Rome].
\textbullet~ 
A. Chakraborty, G. Mukherjee, S. S. Manna,  \textit{Conservative self-organized extremal model for wealth distribution}, Fractals \textbf{20} 163 (2012) [Kolkata]. \textbullet~ S. Kumar and N. Deo,  \textit{Correlation and network analysis of global financial indices}, Phys. Rev. E \textbf{86} 026101 (2012) [Delhi].  
\textbullet~ 
S. Sinha and B. K. Chakrabarti, \textit{Econophysics: An emerging discipline}, Economic \& Political Weekly \textbf{46} 44 (2012) [Chennai \& Kolkata]. 
\textbullet~ P. Sen,  \textit{Nonconservative kinetic exchange model of opinion dynamics with randomness and bounded confidence}, Phys. Rev. E \textbf{86} 016115 (2012) [Kolkata]. 
\textbullet~
A. K. Chandra, \textit{Percolation in a kinetic opinion exchange model}, Phys. Rev. E \textbf{85} 021149 (2012) [Kolkata]. 
\textbullet~
S. Biswas, A. Chatterjee and P. Sen, \textit{Disorder induced phase transition in kinetic models of opinion dynamics}, Physica A  \textbf{391} 3257 (2012) [Kolkata \& Marseille]. 
\textbullet~
R. Rahaman, P. Majumdar and B. Basu,\textit{ Quantum Cournot equilibrium for the Hotelling--Smithies model of product choice}, J. Phys. A \textbf{45} 455301 (2012) [Bergen \& Kolkata].
\vskip 0.2cm 
\noindent
\textbf{Other publications  by Indian scientists in some  Indian journals}: \\
\noindent \textbf{2008}
\textbullet~Bikas K. Chakrabarti, Arnab Chatterjee, Pratip Bhattacharyya, \textit{Two-fractal overlap time series: Earthquakes and market crashes}, Pramana \textbf{71} 203 (2008); Proc. Statphys IITG 2008 [Kolkata]. \\
\noindent \textbf{2009}
\textbullet~ S. Sinha and B. K. Chakrabarti, \textit{Towards a physics of economics}, Physics News \textbf{39} (no. 2) (2009) 33-46 [Kolkata \& Chennai].\\
\noindent \textbf{2010}
\textbullet~S . Sinha, \textit{Are large complex economic systems unstable?} Science and Culture (Special Issue on
Econophysics) \textbf{76} 454-458 (2010) [Chennai]. \textbullet~ B. K. Chakrabarti and A. Chatterjee \textit{The Story of Econophysics}, Science and Culture (Special Issue on Econophysics) \textbf{76}  296-304 (2010) [Kolkata \& Trieste]. \textbullet~M. Lallouache, A. Chakraborti and  B. K. Chakrabarti, \textit{Kinetic exchange models for social opinion formation}, Science and Culture (Special Issue on Econophysics) \textbf{76}  296-304 (2010) [France \& Kolkata]. \newline 
[Other contributions in this Special Issue of Science and Culture \cite{sci-cul}: France- 5, Italy- 5, Japan- 3, USA- 3, Switzerland- 2, Argentina-1, Belgium-1, Brazil-1, China-1, Estonia-1, Finland -1, Germany-1, Ireland-1, Poland-1, Portugal-1, Spain-1 and  UK-1.]

\section{Edited books and conference proceedings volumes from India}
    
Edited Book \cite{ebook-Chakrabarti}:
\begin{itemize}
 \item Eds. B. K. Chakrabarti, A. Chakraborti and A. Chatterjee, \textit{Econophysics and Sociophysics: Trends and Perspectives}, Wiley-VCH, Berlin, 2006.
\end{itemize}

\vskip 0.3cm

\noindent List of proceedings volumes \cite{proc}:
\begin{itemize}
\item  Eds. A. Chatterjee, S. Yarlagadda, B. K. Chakrabarti, \textit{Econophysics of Wealth Distributions},  New Economic Windows, Springer-Verlag, Milan, 2005
[proceedings of ECONOPHYS-KOLKATA I: Econophysics of Wealth Distributions, 15-19 March 2005; Organized by Saha Institute of Nuclear Physics].

\item Eds. A. Chatterjee, B. K. Chakrabarti, \textit{Econophysics of Stock and other Markets},  New Economic Windows, Springer-Verlag, Milan, 2006 [proceedings of ECONOPHYS-KOLKATA II: Econophysics of Stock Markets and Minority Games, 14-17 February 2006;  Organized by Saha Institute of Nuclear Physics].

\item Eds. A. Chatterjee, B. K. Chakrabarti, \textit{Econophysics of Markets and Business Networks}, New Economic Windows, Springer-Verlag, Milan, 2007 [proceedings of ECONOPHYS-KOLKATA III: Econophysics \& Sociophysics of Markets and Networks, 12-15 March 2007; Organized by Saha Institute of Nuclear Physics].

\item Eds. B. Basu, B. K. Chakrabarti, S. R. Chakravarty, K. Gangopadhyay, \textit{Econophysics \& Economics of Games, Social Choices and Quantitative Techniques},  New Economic Windows, Springer-Verlag, Milan, 2010 [proceedings of ECONOPHYS-KOLKATA IV : Econophysics of Games and Social Choices, 9-13 March 2009; Jointly organized by Saha Institute of Nuclear Physics, Kolkata \& Indian Statistical Institute, Kolkata].

\item Eds. F. Abergel, B. K. Chakrabarti, A. Chakraborti, Manipushpak Mitra, \textit{Econophysics of Order-driven Markets},  New Economic Windows, Springer-Verlag, Milan, 2011 [proceedings of ECONOPHYS-KOLKATA V : Econophysics of Order-Driven Markets, 9-13 March 2010; Jointly organized by Saha Institute of Nuclear Physics, Indian Statistical Institute and Ecole Centrale Paris ].

\item  Eds. F. Abergel, B. K. Chakrabarti, A. Chakraborti, A. Ghosh, \textit{Econophysics of Systemic Risk and Network Dynamics}, New Economic Windows, Springer-Verlag, Milan, 2012 [proceedings of ECONOPHYS-KOLKATA VI : Econophysics of Systemic Risk and Network Dynamics, 21-25 October 2011; Jointly organized by Saha Institute of Nuclear Physics and Ecole Centrale Paris].

\item Eds. F. Abergel, H. Aoyama, B.K. Chakrabarti, A. Chakraborti and A. Ghosh,\textit{Econophysics of Agent-based models}, to be published by Springer International Publishing Switzerland, 2013 [proceedings of ECONOPHYS-KOLKATA VII : Econophysics of Agent-based models, 8-12 November 2012; Jointly organized by Saha Institute of Nuclear Physics, Ecole Centrale Paris and Kyoto University].
\end{itemize}

\vskip 0.2cm

\noindent Papers in  \textbf{Econophysics and Sociophysics} by Indian scientists  are:
\newline
\textbullet~ A. K. Gupta, \textit{Models of Wealth Distributions - A Perspective}, pp 161-189 [Panskura].
\textbullet~ S. Sinha and R. K. Pan, \textit{How a ``Hit'' is Born: The Emergence of Popularity from the
Dynamics of Collective Choice}, pp 417-448 [Chennai].
\textbullet~ P. Sen,  \textit{Complexities of Social Networks: A Physicist's Perspective}, pp 473-506 [Kolkata].
\textbullet~ S. Jain and S. Krishna,  \textit{Can we Recognize an Innovation?: Perspective from an Evolving Network Model}, pp 561-591 [Delhi].

Other contributions from: Germany-5, Japan-4, France-3, UK-2,  Belgium-1, Chain-1, Denmark-1, USA-1. 

\vskip 0.4cm
\noindent The list of papers published in the  above proceedings volumes written  by Indian scientists:
\vskip 0.3cm
\noindent Papers in \textbf{Econophysics of Wealth Distributions:}
\textbullet~  S. Sinha, R. K. Pan, Blockbusters,\textit{ Bombs and Sleepers: The income distribution of movies}, pp 43-47 [Chennai]. 
\textbullet~  A. Chatterjee and B. K Chakrabarti, \textit{Ideal-Gas Like Markets: Effect of Savings}, pp 79-92 [Kolkata].
\textbullet~ K. Bhattacharya, G. Mukherjee and S. S.  Manna, \textit{Detailed simulation results for some wealth distribution models in econophysics}, pp 111-119 [Kolkata].
\textbullet~  S.  Yarlagadda and Arnab Das,  \textit{A Stochastic Trading Model of Wealth Distribution}, pp 137-148 [Kolkata]. 
\textbullet~ S.  Sinha, \textit{The rich are different! Pareto law from asymmetric interactions in asset exchange models}, pp 177-183 [Chennai].
\textbullet~ I. Bose and  S. Banerjee, \textit{ A stochastic model of wealth distribution}, pp 195-198 [Kolkata].
\textbullet~ A. Mehta, A. S. Majumdar and  J. M.  Luck, \textit{How the Rich Get Richer}, pp 199-204 [Kolkata \& France].
\textbullet~ D. Bagchi, \textit{Power-Law Distribution in an Emerging Capital Market},  pp 205-209 [Kolkata].  
\textbullet~ A. Sarkar and  P. Barat, \textit{Statistical Analysis on Bombay Stock Market}, pp 210-213 [Kolkata].
\textbullet~  D.  P.  Pal and  H.  K. Pal,  \textit{Income Distribution in the Boltzmann-Pareto Framework}, pp 218-222 [Kolkata].
\textbullet~ B. K. Chakrabarti, \textit{Econophys-Kolkata: A Short Story}, pp 225-228 [Kolkata]. 

Other contributions from: Germany-5, Japan-4, Argentina-2, Italy-2, UK-2, USA-2,  Brazil-1, China-1, Finland-1, France-1. 

\vskip 0.4cm
\noindent Papers in  \textbf{Econophysics of Stock and other Markets:}
\textbullet~ A. Chakraborti,  \textit{An Outlook on Correlations in Stock Prices}, pp 13-23 [Kolkata].
\textbullet~ S. Sinha and  R. K. Pan, \textit{The Power (Law) of Indian Markets: Analysing NSE and BSE Trading Statistics},  pp 24-34 [Chennai].
\textbullet~ V. Kulkarni and  N. Deo, \textit{A Random Matrix Approach To Volatility In An Indian Financial Market}, pp 35-48 [Delhi].
\textbullet~  A. Sarkar and P. Barat,  \textit{Fluctuation Dynamics of Exchange Rates on Indian Financial Market} [Kolkata], pp 67-76.
\textbullet~  D.  Bagchi, \textit{ Noise Trading in an Emerging Market: Evidence and Analysis}, pp 77-84 [Kolkata].
\textbullet~ U.  K. Basu,  \textit{How Random is the Walk: Efficiency of Indian Stock and Futures Markets}, pp 85-97 [Kolkata].
\textbullet~  B.  K. Chakrabarti, A. Chatterjee and P.  Bhattacharyya, \textit{Two Fractal Overlap Time Series and Anticipation of Market Crashes}, pp 153-158 [Kolkata].
\textbullet~ S.  Sinha, \textit{The Apparent Madness of Crowds: Irrational Collective Behavior Emerging from Interactions among Rational Agents}, pp 158-162 [Chennai].
\textbullet~ M.  Mitra,  \textit{Information Extraction in Scheduling Problems with Non-Identical Machines}, pp 175-182 [Kolkata].
\textbullet~   P. Manimaran, J. C. Parikh, P. K. Panigrahi S. Basu, C. M. Kishtawal and  M. B. Porecha,\textit{ Modelling Financial Time Series}, pp 183-191 [Kanpur].
\textbullet~  M. S. Santhanam, \textit{Random Matrix Approach to Fluctuations and Scaling in Complex Systems}, pp 192-200 [Ahmedabad].
\textbullet~ A.  Sarkar, \textit{Regional Inequality}, pp 208-218 [Kolkata].
\textbullet~ B. K. Chakrabarti, \textit{A Brief History of Economics: An Outsider`s Account}, pp 219-224 [Kolkata].

Other Contributions from: China-3, UK-3, Italy-2, Japan-2, USA-2, Finland-1, France-1,  Hungary-1, Ireland-1.  

\vskip 0.4cm
\noindent Papers in \textbf{Econophysics of Markets and Business Networks}
\textbullet~  S. Sinha and  R. K. Pan,  \textit{Uncovering the Internal Structure of the Indian Financial Market: Large Cross-correlation Behavior in the NSE}, pp 3-20 [Chennai].
\textbullet~  A. Chakraborti, M. Patriarca and  M. S. Santhanam,\textit{ Financial Time-series Analysis: a Brief Overview}, pp 51-68 [Banaras, Estonia \& Ahmedabad].
\textbullet~ K. B. K. Mayya and M. S. Santhanam, \textit{Correlations, Delays and Financial Time Series}, pp 69-76 [Ahmedabad].
\textbullet~  K. Bhattacharya, G. Mukherjee and S. S. Manna,\textit{ The International Trade Network}, pp 139-148 [Kolkata].
\textbullet~ S. Sinha and N. Srivastava,  \textit{Is Inequality Inevitable in Society? Income Distribution as a Consequence of Resource Flow in Hierarchical Organizations}, pp 215-226 [Kolkata].
\textbullet~  A. Sarkar,  \textit{Knowledge Sharing and R\&D Investment}, pp 227-232 [Kolkata].
\textbullet~  M.  Mitra, \textit{Preferences Lower Bound in the Queueing Model}, pp 233-238 [Kolkata].
\textbullet~  B. K. Chakrabarti,  \textit{Kolkata Restaurant Problem as a Generalised El Farol Bar Problem}, pp 239-246 [Kolkata].

Other contributions from: Italy-3, China-2,  Japan-2, UK-2, Canada-1, Estonia-1,  Poland-1, USA-1.  

\vskip 0.4cm
\noindent  Papers in  \textbf{Econophysics and Economics of Games, Social Choices and Quantitative Techniques}
\textbullet~  A.  Ghosh, A.  S. Chakrabarti, and B.  K. Chakrabarti, \textit{ Kolkata Paise Restaurant Problem in Some Uniform Learning Strategy Limits}, pp 3-9 [Kolkata].
\textbullet~  D.  Mishra and M.  Mitra, \textit{ Cycle Monotonicity in Scheduling Models}, pp 10-16 [Delhi \& Kolkata].
\textbullet~ S.  Yarlagadda,  \textit{Using Many-Body Entanglement for Coordinated Action in Game
Theory Problems}, pp 44-51 [Kolkata].
\textbullet~  K.  Gangopadhyay and B.  Basu,  \textit{The Morphology of Urban Agglomerations for Developing Countries: A Case Study with China}, pp 90-97 [Kolkata].
\textbullet~  V.  S. Vijayaraghavan and S.  Sinha, \textit{A Mean-Field Model of Financial Markets: Reproducing Long Tailed Distributions and Volatility Correlations}, pp 98-109 [Chennai].
\textbullet~  P.  K. Panigrahi, S. Ghosh, P. Manimaran, and D. P. Ahalpara, \textit{Statistical Properties of Fluctuations: A Method to Check Market Behavior}, pp 110-118 [Ahmedabad, Chennai, Hyderabad \& Gandhinagar].
\textbullet~ A. K. Ray, \textit{Modeling Saturation in Industrial Growth}, pp 119-124 [Mumbai].
\textbullet~ V. A. Singh, P. Pathak and P. Pande, \textit{ Monitoring the Teaching - Learning Process via an Entropy Based Index}, 139-146 [Mumbai \& Kanpur].
\textbullet~ J. Basu, B. Sarkar, and A. Bhattacharya,  \textit{Technology Level in the Industrial Supply Chain: Thermodynamic Concept}, pp 147-153 [Kolkata].
\textbullet~ A. Sarkar, S. Sinha, B. K. Chakrabarti, A. M. Tishin, and V.I.
Zverev,  \textit{Discussions and Comments in Econophys Kolkata IV},  pp 154-171 [Kolkata, Chennai \& Moscow].
\textbullet~  S. Subramanian,\textit{ Variable Populations and Inequality-Sensitive Ethical Judgments}, pp 181-191 [Chennai].
\textbullet~ S. R. Chakravarty and S. Ghosh, \textit{A Model of Income Distribution}, pp 192-203 [Kolkata].
\textbullet~ V. K. Ramachandran, M. Swaminathan, and A. Bakshi, \textit{Food Security and Crop Diversification: CanWest Bengal Achieve Both?}, pp 233-240 [Kolkata].
\textbullet~ A. Majumder and M. Chakrabarty, \textit{Estimating Equivalence Scales Through Engel Curve Analysis}, pp 241-251 [Kolkata].
\textbullet~ S. Das and M. Chakrabarty, \textit{Testing for Absolute Convergence: A Panel Data Approach}, pp 252-262 [Kolkata].
\textbullet~  S.  Datta and A. Mukherji, \textit{Goodwin's Growth Cycles: A Reconsideration}, pp 263-276 [Delhi].
\textbullet~  B. Chakraborty and M. R. Gupta, \textit{Human Capital Accumulation, Economic Growth and Educational
Subsidy Policy in a Dual Economy}, pp 277-292 [Kolkata].
\textbullet~  B.  S.  Chakraborty and A. Sarkar, \textit{Trade andWage Inequality with Endogenous Skill Formation}, pp 306-319 [Kolkata].
\textbullet~  M. Mitra and A.  Sen,  \textit{Dominant Strategy Implementation in Multi-unit Allocation Problems}, pp 320-330 [Kolkata \& Delhi].
\textbullet~ A.  Kar, \textit{Allocation through Reduction on Minimum Cost Spanning Tree Games}, pp 331-346 [Delhi].
\textbullet~  A. Kar, M.  Mitra and S.  Mutuswami, \textit{A Characterization Result on the Coincidence of the Prenucleolus and the Shapley Value}, pp 362-371 [Delhi, Kolkata \& Leicester].
\textbullet~  K. Ghosh Dastidar, \textit{Reflecting on Market Size and Entry under Oligopoly}, pp 381-394 [Delhi].

Other contributions from: USA-7, Italy-3, UK-2,  Canada-1,  Hungary-1, Japan-1, Poland-1, Russia-1.

\vskip 0.4cm
\noindent Papers in \textbf{Econophysics of Order-driven Markets}
\textbullet~  V. S. Vijayaraghavan and  S. Sinha, \textit{Are the Trading Volume and the Number of Trades Distributions Universal}? pp 17-30 [Chennai].
\textbullet~ S. R. Chakravarty, D. Chakrabarti,  \textit{The von Neumann–Morgenstern Utility Functions with Constant Risk Aversions}, pp 253-258 [Kolkata \& Panskura].
\textbullet~  K. Gangopadhyay and B. Basu, \textit{Income and Expenditure Distribution. A Comparative Analysis}, pp 259-270 [Kozhikode \& Kolkata].
\textbullet~ M. Mitra, \textit{Two Agent Allocation Problems and the First Best}, pp 271-276 [Kolkata].
\textbullet~  A. Chakraborti, B.  K. Chakrabarti, \textit{Opinion Formation in the Kinetic Exchange Models}, pp 289-304 [Paris \& Kolkata].

Other Contributions from: France-11, USA-4, Germany-2, Italy-2, Austria-1.

\vskip 0.4cm
\noindent  Papers in \textbf{Econophysics of Systemic Risk and Network Dynamics}
\textbullet~ S. Sinha, M. Thess, and S. Markose, \textit{How Unstable Are Complex Financial Systems? Analyzing an Inter-bank Network of Credit Relations}, pp 59-76 [Chennai, Germany \& UK].
\textbullet~  K. Gangopadhyay and B. Basu, \textit{Evolution of Zipf`s Law for Indian Urban Agglomerations Vis-\`{a}-Vis Chinese Urban Agglomerations}, pp 119-132 [Kozhikode \& Kolkata] .
\textbullet~  A. Mehta, \textit{Predatory Trading and Risk Minimisation: How to (B)Eat the Competition}, pp 141-156 [Kolkata].
\textbullet~ A. Ghosh, S. Biswas, A. Chatterjee, A. S. Chakrabarti, T. Naskar, M.  Mitra, and B. K. Chakrabarti, \textit{Kolkata Paise Restaurant Problem: An Introduction}, pp  173-200 [Kolkata \& Espoo].
\textbullet~  P.  Banerjee, M. Mitra, and C.  Mukherjee, \textit{Kolkata Paise Restaurant Problem and the Cyclically Fair Norm,} pp 201-216 [Kolkata].
\textbullet~ S. Kumar and N. Deo,  \textit{Analyzing Crisis in Global Financial Indices}, pp 261-276 [Delhi].
\textbullet~  K. C. Dash and M. Dash, \textit{Study of Systemic Risk Involved in Mutual Funds}, pp 277-286 [Rourkela \& Pune].
\textbullet~  P. K. Panigrahi, S. Ghosh, A. Banerjee, J. Bahadur, and P. Manimaran, \textit{Characterizing Price Index Behavior Through Fluctuation Dynamics}, pp 287-295 [Kolkata, Durban \& Hyderabad].

Other Contributions from:  France-4,  Japan-3, Italy-2,  Switzerland-2, Finland-1, Germany-1, UK-1,  Hungary-1, Sweden-1.

\vskip 0.4cm
\noindent  Papers in \textbf{Econophysics of Agent-based models}
\textbullet~  K. Gangopadhyay and K. Guhathakurta, \textit{Agent based modeling of Housing asset bubble: A simple utility function based investigation} [Kozhikode].
\textbullet~  A. Ghosh, A. S. Chakrabarti, A. K. Chandra and A. Chakraborti, \textit{An (n+1)-th look at kinetic exchange models} [Kolkata, Boston \& Paris].
\textbullet~  K. R. Chowdhury, A. Ghosh, S. Biswas and B. K. Chakrabarti,  \textit{Kinetic exchange opinion model: solution in the single parameter map limit} [Kolkata].
\textbullet~ S. Sinha and U. Kovur, \textit{Uncovering the network structure of the world currency market:
Cross-correlations in the fluctuations of daily exchange rates} [Chennai \& Pilani].
\textbullet~  K. C. Dash, \textit{Evolution of Econophysics} [Rourkela].
\textbullet~ A. Ghosh and A. S. Chakrabarti, \textit{Econophysics and sociophysics: Problems and prospects} [Kolkata \& Boston].

Other contributions from: Japan-5, France-3, USA-2, Argentina-1, Belgium-1, Italy-1,  Netherlands-1, Switzerland-1. 
\vskip 0.2 cm
\noindent \textbf{Papers by Indian scientists  in other proceedings volumes:}\\
\textbf{2004} \textbullet~  B. K. Chakrabarti and  A. Chatterjee, \textit{Ideal Gas-Like Distributions in Economics: Effects of Saving Propensity}, in `Applications of Econophysics', Ed. H. Takayasu, pages 280-285 (2004), Conference proceedings of Second Nikkei Symposium on Econophysics, Tokyo, Japan, 2002, by Springer-Verlag, Tokyo [Kolkata]. \\
\textbf{2005}
\textbullet~  A. Chatterjee, B. K. Chakrabarti, R. B. Stinchcombe, \textit{Analyzing money distributions in `ideal gas' models of markets} in `Practical Fruits of Econophysics', Ed. H. Takayasu, pages 333-338 (2005), Springer-Verlag, Tokyo; Conference proceedings of Third Nikkei Symposium on Econophysics, Tokyo, Japan, 2004 [Kolkata \& UK].
\textbullet~B. K. Chakrabarti, A. Chatterjee, P. Bhattacharyya, \textit{Time series of stock price and of two fractal overlap: Anticipating market crashes?} in `Practical Fruits of Econophysics', Ed. H. Takayasu, pages 107-110 (2005), Springer-Verlag, Tokyo; Conference proceedings of Third Nikkei Symposium on Econophysics, Tokyo, Japan, 2004 [Kolkata]. \\
\textbf{2006}
\textbullet~  S. Sinha and S. Raghavendra, \textit{Market polarization in presence of individual choice volatility}, Advances in Artificial Economics: The Economy as a Complex Dynamic System (Ed. C. Bruun) Springer,  177 (2006) [Chennai]. \textbullet~ S. Sinha and S. Raghavendra, \textit{Emergence of two-phase behavior in markets through interaction and learning in agents with bounded rationality}, Practical Fruits of Econophysics (Ed. H. Takayasu) Springer, 200 (2006) [Chennai].\\
\textbf{2012}
\textbullet~ K. Guhathakurta, S. Bhattacharya, S. Banerjee and  B. Bhattacharya, \textit{Examining the relative Nonlinear dynamics of stock and commodity indices in emerging and developed market} in S. Banerjee, Chaos \& Complexity Theory for Management (pp. 63-88). IGI Global (2012) [Kolkata \& Kozhikode].

\begin{table}

\begin{tabular}{cc}
     \begin{minipage}{.6\linewidth}
    \begin{tabular}{|p{4cm}|l|}
    \hline
   Name of the City & No. of Papers  \\ \hline
    Kolkata &  97 \\ \hline
    Chennai  &  26 \\ \hline
    Delhi  &  12  \\ \hline 
    Ahmedabad/Gandhinagar  &  9  \\ \hline 
    Kozhikode  &  5  \\ \hline
    Panskura \begin{tiny}(East Midnapore)\end{tiny}  &  4  \\ \hline 
    Hyderabad  &  3  \\ \hline 
    Mumbai  &  3  \\ \hline
    Pune  &  3  \\ \hline 
    Kanpur  &  2  \\ \hline 
    Rourkela  &  2  \\ \hline 
    Banaras  &  1  \\ \hline
    Pilani  &  1  \\ \hline

    \hline
    \end{tabular}\end{minipage} &
\begin{minipage}{.4\linewidth}
         \begin{tabular}{|p{1cm}|p{1cm}|}
    \hline
  Item  & Total Count  \\ \hline
  Papers & 148 \\ \hline
  Books  & 4  \\  \hline 
  Conf.  Proc. Vol. & 7 \\ \hline
  Edited Books  & 1 \\ \hline
   
    \end{tabular}

\end{minipage}

\end{tabular}

\caption{(Left) Numbers of econophysics papers (in journals and conf. proc. vol. given in sections 4 and 5) by Indian scientists. (Right) Total  contribution  of papers, books, edited books and conference proceedings volumes from India (published internationally).}
\end{table}
\section{Institutions where the researches have so far been  carried out}
The researches  on interdisciplinary research fields in India is not any recent trend. In 1931  Prasanta Chandra Mahalanobis established  the Statistical Laboratory at Kolkata. Later the institution was named to  Indian Statistical Institute. The main motivation of the institution was  research and training of Statistics, development of theoretical Statistics and its applications in various natural and social sciences.  In the  last two decades the major developments on  such interdisciplinary research (on econophysics or sociophysics) have  come from Saha Institution Nuclear Physics. This is also the  place where the term `econophysics' was coined in 1995. A large number of  papers on econophysics   have been published from  this institution and significant research activities are also being continued. A major international conference series  on econophysics, namely `Econophys-Kolkata'  is being  organized regularly  here (seven events since 2005; see section 5). Other places where econophysics research is being  actively pursued  are:   universities like Calcutta University, Delhi University, Pune University, etc. and institutions like  Institute of Mathematical Sciences (Chennai),  S N Bose National Centre for Basic Science (Kolkata), Tata Institute of Fundamental Research (Mumbai), Indian Institute of Management (Kozhikode), etc.  National level conferences on econophysics are now being held in several places; in particular,  Institute of Mathematical Sciences  holds them quite regularly since 2004: First one was ``Workshop on The Economy as a Complex System'', Dec 6-7, 2004, the second one was discussion meeting on ``The Economy as a Complex System II: Economic Dynamics'', Dec 27-29, 2010 and the third one was ``Brainstorming Meeting on Econophysics: Science for the Economy'', July 30, 2013.\footnote{Other (University Grant Commission sponsored) national level conferences on econophysics include ``Physics of Financial Markets - Challenges and Opportunities'', Sept. 17-18, 2011, at Neelashaila Mahavidyalaya, Rourkela (Sambalpur University, Odisha) and ``Econophysics'', Aug. 18-19, 2012, at Hindol College, Khajuriakata (Utkal University, Odisha).}
\newline \newline
We give below a graphical presentation  of the Indian cities where the major econophysics researches have been  carried out  so far (data taken  from sections 4 and 5, Tables 1 and  2).
\begin{figure}[!h]
\centering
\includegraphics[height=6cm]{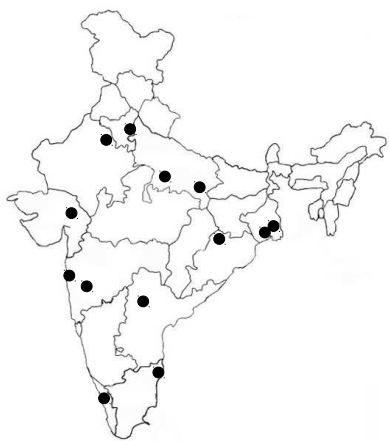}
\caption{The dots represent the locations where  major researches  on econophysics  have so far  been carried out (see Table 1).}
\label{fig3}
\end{figure}

\begin{figure}[!h]
\centering
\includegraphics[height=7cm]{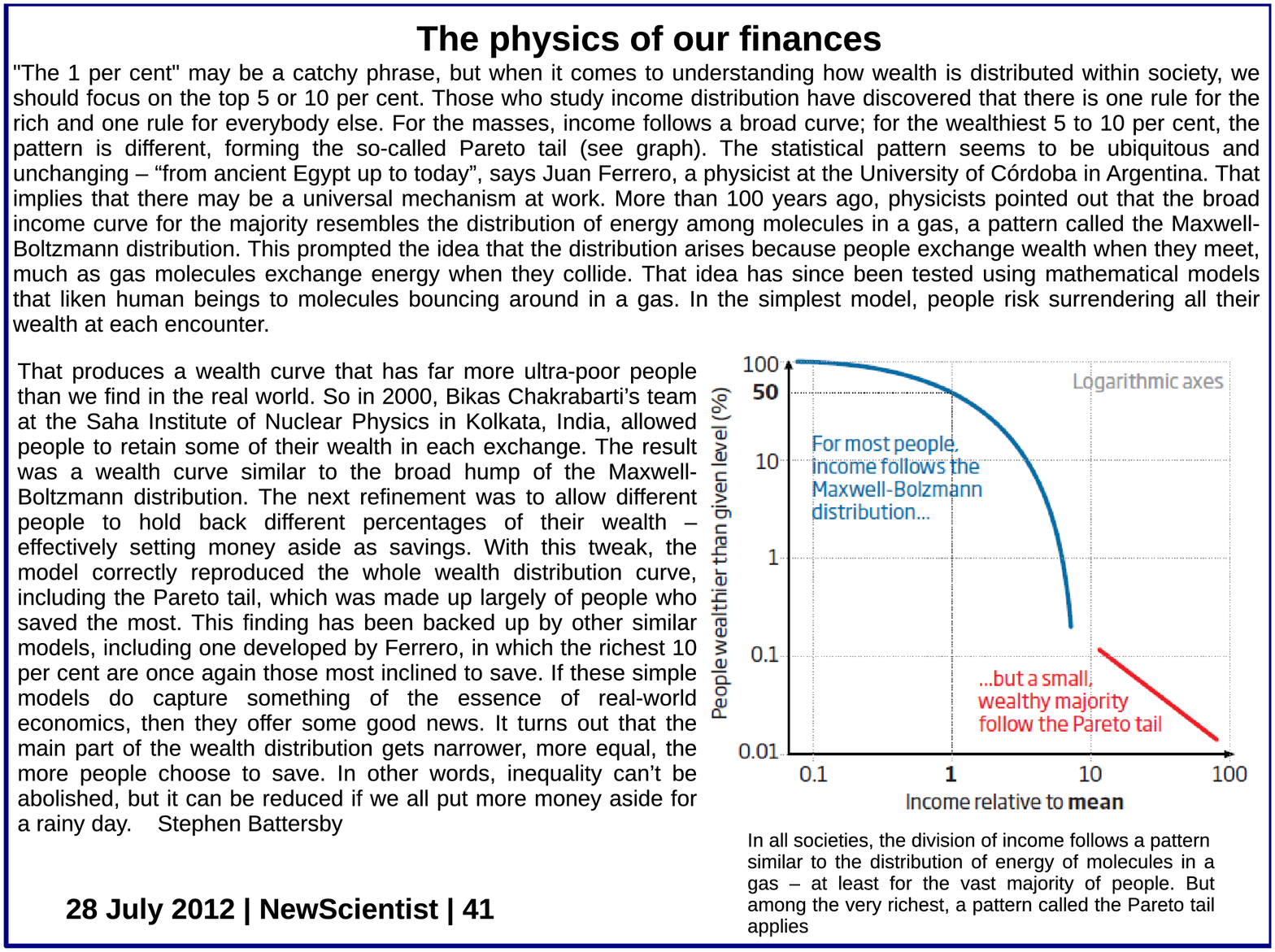}
\caption{One of the recent  reporting  on  econophysics research from India.}
\label{fig4}
\end{figure}
\section{Impact of Indian researches}
\begin{table}
\begin{center}
 
     \begin{tabular}{|p{1.8cm}|l|p{1.8cm}|l|}
    \hline
    Country name  &  Papers &  Country name  &   Papers \\ \hline
    India &  73 &  Switzerland & 3 \\ \hline
    France   &  20 &  Canada & 2   \\ \hline
    USA   &  18  &  Poland & 2   \\ \hline 
    Japan   &  17 &  Austria & 1 \\ \hline
    Italy & 15 & Belgium & 1  \\ \hline
    UK & 10 & Brazil & 1\\ \hline
    Germany & 8 & Estonia & 1  \\ \hline
    China & 6 & Ireland & 1 \\ \hline
    Argentina & 3 &  Netherlands& 1 \\ \hline
    Finland & 3 & Russia  & 1 \\ \hline
    Hungary & 3 & Sweden & 1 \\ \hline
       \hline
    \end{tabular}
   \caption{ Number of papers contributed by the researchers   of  different countries in the proceedings volumes of  \textit{Econophys-Kolkata} conference series (I-VII so far; section 5). }
\end{center}
\end{table}

The  term  `econophysics' has now been included \cite{Dictionary}  in `The New Palgrave Dictionary of Economics'. The entry (written by economist  J. Barkley Rosser Jr) starts with  ``According to Bikas Chakrabarti (\dots), the term econophysics was neologized in 1995 at the second Statphys-Kolkata conference in Kolkata (formerly Calcutta), India  \dots ''.   Another note \cite{Battersby}  ``The physics of our finances'' published in  New Scientist  in July 2012    highlighted the contributions from India (see Fig. \ref{fig4}). Recently  an entry on  econophysics has also  been  included  in  \textit{``Encyclopedia of Philosophy and the Social Sciences''} published by SAGE Publications (2013) and the entry on it has been written  by Bikas K. Chakrabarti \cite{sage}.

The International conference series `Econophys-Kolkata' had started in 2005 and have already seven events held in Kolkata. It is now jointly  sponsored by   Saha Institute of Nuclear Physics, Ecole Centrale Paris and Kyoto University. Also, the contributions by foreign researchers  in the proceedings volumes (seven so far: Proc. volumes have all  been  published in New Economic Windows, Springer-Verlag,  given in section 5, Table 2) indicate the impact of Indian researches in econophysics and  sociophysics internationally.

\begin{figure}[!h]
\centering
\includegraphics[height=7cm]{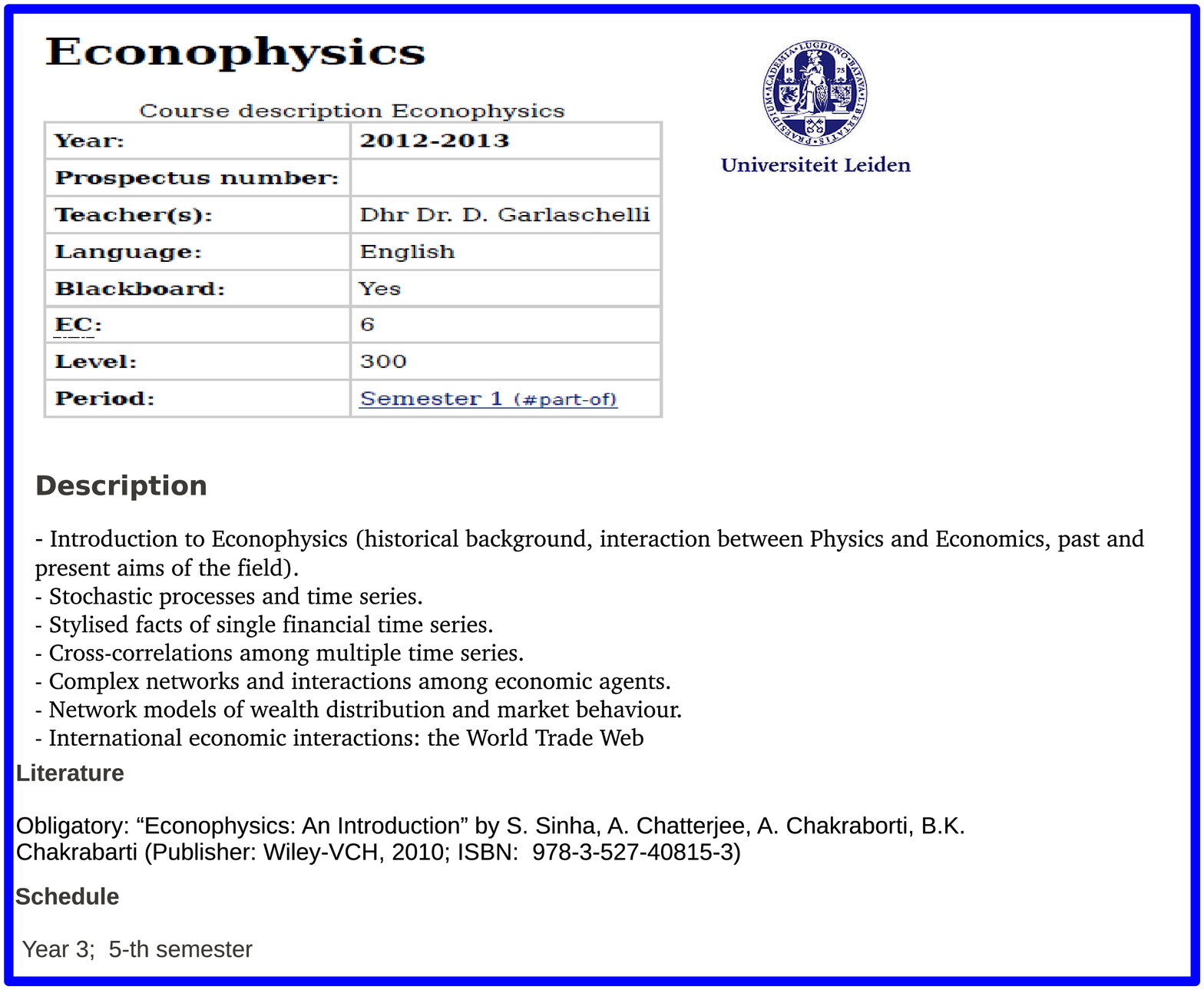}
\caption{Reproduced  from web page of the Econophysics course  in   Leiden University. (https://studiegids.leidenuniv.nl/en/courses/show/34804/econofysica). For their econophysics courses in the last three years,  ``Econophysics: An Introduction'', a book from India, is being followed (obligatory literature).}
\label{fig5}
\end{figure}

The first text book (in physics) on econophysics entitled  ``\textit{Econophysics: An Introduction}''  has been  written by Indian scientists (see section 3).  This book is  already being  followed by many universities  outside India for their graduate courses (see Fig. \ref{fig5} for the  econophysics course in Leiden university).  In fact  among the formal courses on econophysics,  the one offered by the Physics Department of the Leiden University is particularly noteworthy:  From this department,   the first Nobel-laureate (in 1969) in economics Jan Tinbergen came. Also here the first professor chair for econophysics had been created in 2010.

\section{Concluding remarks}
Here we have given the  statistics on the development of econophysics  by Indian researchers.  Many researchers from  India have been involved  in econophysics research from the formal  beginning of the subject in 1995, and many Indian research institutes and universities are involved in this research area (see table I). Apart from  publications of important papers (section 4), several important  conference proceedings volumes and edited volumes (section 5), research monographs  and   text books (section 3) on econophysics have been published from India. Some of these  papers have made good impact and some of these books are being widely used in econophysics and sociophysics courses started in many well known universities in Europe and elsewhere (see e.g., Fig. \ref{fig5}).  There have also been some attempts  to initiate such formal research groups or centers   in India. In particular, the  ``Policy Planing \& Evaluation Committee'' (PPEC) of the   Indian Statistical Institute,  in its June 22 (2011) meeting considered  a ``Proposal for building a Center for Econophysics \& Quantitative Finance Research'' and recommended that ``PPEC recognizes this to be an important proposal, but considering the availability of manpower and the current focus of ERU (Economic Research Unit), it recommends that the proposal be carried as a plan research project, but not as a full-fledged center at this point of time. However, the recruitment of faculty members in the area of econophysics or related disciplines may be made in ERU if need.''\footnote{Private communication:  Satya Ranjan Chakravarty (Indian Statistical Institute) and Bikas K. Chakrabarti (Saha Institute of Nuclear Physics)}  We are also happy to learn  that   similar endeavors are  being  made  in other important institutions of the country.

\end{document}